%% file: paper.tex
\documentclass[aps,prc,groupedaddress,superscriptaddress,amsmath,floatfix,twocolumn]{revtex4}

\usepackage{nicefrac}
\usepackage{pst-all}
\usepackage{graphicx}

\newpsobject{PST@cascd}{psdots}{linecolor=red,      linewidth=.02,  linestyle=solid,dotsize=.05pt 1,dotstyle=x}
\newpsobject{PST@cascl}{psline}{linecolor=red,      linewidth=.0025,linestyle=solid}
\newpsobject{PST@hgeos}{psline}{linecolor=blue,     linewidth=.0025,linestyle=solid}
\newpsobject{PST@xieos}{psline}{linecolor=orange,   linewidth=.0035,linestyle=dashed,dash=.01 .01}
\newpsobject{PST@bmeos}{psline}{linecolor=violett,  linewidth=.0050,linestyle=dotted,dotsep=.002}

\newpsobject{PST@inc}{psline}{linecolor=black,    linewidth=.0025,linestyle=solid}
\newpsobject{PST@aft}{psline}{linecolor=red,      linewidth=.0025,linestyle=dashed,dash=.01 .01}
\newpsobject{PST@hyd}{psline}{linecolor=blue,     linewidth=.0035,linestyle=dotted,dotsep=.004}
\newpsobject{PST@bef}{psline}{linecolor=darkgreen,linewidth=.0025,linestyle=dashed,dash=.01 .004 .004 .004}

\newpsobject{PST@inklusiv}{psline}{linecolor=black,    linewidth=.0025,linestyle=solid}
\newpsobject{PST@pirhogpi}{psline}{linecolor=red,      linewidth=.0025,linestyle=dashed,dash=.01 .01}
\newpsobject{PST@pipigrho}{psline}{linecolor=blue,     linewidth=.0035,linestyle=dotted,dotsep=.004}
\newpsobject{PST@etachann}{psline}{linecolor=darkgreen,linewidth=.0025,linestyle=dashed,dash=.01 .004 .004 .004}
\newpsobject{PST@gammagam}{psline}{linecolor=orange,   linewidth=.0025,linestyle=dashed,dash=.02 .01}
\newpsobject{PST@quglupla}{psline}{linecolor=violett,  linewidth=.0035,linestyle=dotted,dotsep=.004}

\newpsobject{PST@bgbox}{psframe}{linecolor=black,linewidth=.001,linestyle=solid,fillcolor=white,fillstyle=solid}

\definecolor{darkgreen}{rgb}{0,.5,0}
\definecolor{violett}{rgb}{.5,0,.5}
\definecolor{orange}{rgb}{.8,.4,0}
\begin{document}
\title{Transport and hydrodynamic calculations of direct photons at FAIR}
\author{Bj\o{}rn B\"auchle}
\email{baeuchle@th.physik.uni-frankfurt.de}
\author{Marcus Bleicher}
\affiliation{Frankfurt Institute for Advanced Studies, Frankfurt am Main, Germany}
\affiliation{Institut f\"ur Theoretische Physik, Goethe-Universit\"at,
Frankfurt am Main, Germany}

\begin{abstract}

The microscopic transport model UrQMD and a micro+macro hybrid model are
used to calculate direct photon spectra from U+U-collisions at
$E_{\sf lab} = 35$~AGeV as will be measured by the CBM-collaboration at
FAIR. In the hybrid model, the intermediate high-density part of the nuclear
interaction is described with ideal 3+1-dimensional hydrodynamics. Different
equations of state of the matter created in the heavy-ion collisions are
investigated and the resulting spectra of direct photons are predicted. The
emission patterns of direct photons in space and time are discussed.

\end{abstract}

\maketitle

\section{Introduction}\label{sec:intro}

In order to understand the phases of compact stellar objects and phase
transitions in the early universe, a deep understanding of the phase diagram
of strongly interacting matter is needed. The experimental tool to gain
these informations is heavy-ion physics, where large nuclei are collided. In
those collisions, nuclear matter is heated to temperatures that are believed
to be sufficient to cause a phase transition from a hadron gas to a state
where partonic degrees of freedom are relevant (the Quark-Gluon-Plasma QGP).
Also at lower energies, but very high baryon densities novel phenomena like
color super conducting phases~\cite{nucl-th/0305030} or even Quarkyonic
Matter~\cite{arXiv:0706.2191} are expected in contrast to the well known
hadron gas.

From lattice gauge theory calculations, it is deduced that the transition
between the hadronic and QGP phase is a
cross-over~\cite{hep-lat/0611014,arXiv:0811.3858}, if the baryochemical
potential $\mu_{\sf B}$ is sufficiently small. The transition temperature is
expected to be around $T_C \approx 170$~MeV~\cite{arXiv:1005.3508}. At high
$\mu_{\sf B}$, first principle calculations of the phase structure are no
longer possible, but symmetry arguments suggest that the phase transition
line ends at high $\mu_{\sf B}$ and zero temperature $T = 0$ as a first
order phase transition. If so, then a critical end point must exist, at
which the cross-over turns into a first order phase transition.

The search for this critical end point is the main motivation for current
and planned experimental heavy-ion programs of the SHINE-experiment at the
Super Proton Synchrotron (CERN-SPS)~\cite{arXiv:0709.1867} and the
CBM-experiment at the Facility for Antiproton and Ion Reserach
(FAIR)~\cite{Hohne:2005qm}. The current low-energy run at the Relativistic
Heavy Ion Collider (BNL-RHIC) also addresses this
issue~\cite{arXiv:0906.0305,arXiv:1007.2613}.

The CBM-collaboration will measure uranium-uranium collisions in a
fixed-target experiment at incident energies of up to $E_{\sf lab} =
35$~AGeV. The proposed features for this detector include, besides
subsystems to measure hadronic observables, electromagnetic calorimetry in
order to measure photons.

Electromagnetic probes such as leptons and photons have the advantage that
they do not suffer from final state interactions and hence carry all
information from their production vertex into the
detector~\cite{arXiv:0904.2184}.  Therefore, they also carry information
from all stages of the heavy-ion collision.  Hadronic probes, on the other
hand, are mostly produced very late in the collision or rescatter in the
late stages, so that they carry only indirect information from the early
stages, such as flow patterns and fluctuations.

Previous calculations of direct photons from transport theory include work
with UrQMD by Dumitru {\it et al.}~\cite{hep-ph/9709487} and B\"auchle {\it
et al.}~\cite{arXiv:0905.4678} and with HSD by Bratkovskaya {\it et
al.}~\cite{arXiv:0806.3465}. Hydrodynamics has been used in many direct
photon calculations, see
e.g.~\cite{Kapusta:1991qp,nucl-th/0006018,Turbide:2003si,hep-ph/0502248,arXiv:0811.0666,arXiv:0902.1303,arXiv:0903.1764,arXiv:0911.2426}.

Most photons produced in heavy-ion collisions, however, come from hadronic
decays, predominantly the decay of the $\pi^0$ and $\eta$. Due to the long
lifetime of both particles, they decay far outside the collision zone. In
order to extract information about the fireball from photons, the decay
photon contribution has to be subtracted. The remaining photons, those that
do not come from decays, are called direct photons. Experimental techniques
for the extraction of direct photon yields from the inclusive measurements
include the direct estimation of the background via invariant mass-analysis
of the photons~\cite{nucl-ex/0006007,nucl-ex/0006008}, the analysis of
interference patterns (using a Hanburry Brown-Twiss
analysis)~\cite{nucl-ex/0310022} and the extrapolation of the spectra of
massive virtual photons (low-mass dileptons) to massless
photons~\cite{arXiv:0912.0244}.

In this work, we investigate the influence of different evolutions of the
bulk medium on the direct photon production in U+U-collisions at $E_{\sf
Lab} = 35$~AGeV using the model established in~\cite{arXiv:0905.4678}. We
compare calculations with hadronic degrees of freedom to calculations with a
first order phase transition to a QGP and calculations with a cross-over to
chirally restored and deconfined matter. In Section~\ref{sec:hybrid}, we
briefly present the model and parameters used for these calculations, and in
Section~\ref{sec:results} we show and analyse our calculations for direct
photon emission.

\section{The model}\label{sec:hybrid}

The microscopic transport model Ultrarelativistic Quantum Molecular Dynamics
(UrQMD)~\cite{Bass:1998ca,Bleicher:1999xi,Petersen:2008kb} provides the
setting in which the present work is calculated. Calculations with UrQMD
using standard options provides a baseline calculation with hadronic and
string degrees of freedom throughout the nucleus-nucleus-interaction and
with vacuum properties of all particles.

Uranium has an eccentricity of $\epsilon=0.27$. However, for the present
study, the uranium nuclei are initialized spherically, which provides an
intrinsic averaging over different possible alignments of the projectile and
target nucleus. We have checked that for untriggered collisions (i.e.\
without selection on the alignment of the nuclei) the results are not
sensitive to this choice.

The hybrid option which is new in version
3.3~\cite{Rischke:1995ir,Rischke:1995mt,Petersen:2008dd,u3.3} allows to
substitute the high-density part of a heavy-ion collision with an ideal
3+1-dimensional hydrodynamic calculation, where different assumptions on the
matter present in the collision can be tested by varying the Equation of
State. Non-equilibrium initial state interactions and final state
scatterings in the dilute hadronic medium, as well as decays are calculated
within the microscopic transport model.

The transition from the particle-based transport description to the
density-based hydrodynamic description happens when the initial
baryon-currents have decoupled. We estimate this to be when the initial
nuclei have completely passed through each other, which in the case of the
system investigated here (U+U-collisions at $E_{\sf lab} = 35$~AGeV)
is at $t_{\sf start} = 3.7$~fm after the very first scatterings. At this
point, all particles that have interacted or are newly produced are used to
calculate energy-density, baryon number-density and momentum-densities,
which are then taken as input for the hydrodynamic calculation. During this
transition, the system is forced into local thermal equilibrium, regardless
of the state of actual equilibration of the matter.

After the hydrodynamic evolution has proceeded, the
Cooper-Frye-formalism~\cite{Cooper:1974mv} is used to couple the
hydrodynamic calculation to the late transport stage. This transition
happens individually for each transverse slice, characterized by the same
position along the beam direction, when all cells in that slice have diluted
to below a critical energy density, whose exact value depends on the
Equation of State employed (see Table~\ref{tab:epscrit}). The hypersurface
on which the transition happens is piecewise isocronous and non-continuous.
However, since after that transition scatterings and decays are calculated
with UrQMD, the effects of that should be negligible.  For more details on
the mappings, the reader is referred to Petersen {\it et
al.}~\cite{Petersen:2008dd,arXiv:0905.3099}. 

\subsection{Equations of State}\label{sec:hybrid:eos}

Four different calculations are compared in the current work. Firstly, we
compare direct photon spectra from hadronic systems, calculated in pure
cascade model and in the cascade+hydrodynamic hybrid model with Hadron-Gas
EoS (HG-EoS)~\cite{nucl-th/0209022}. The HG-EoS has the same degrees of
freedom as the transport phase and therefore provides excellent means to
investigate the effects of instant thermalization at the transition from
transport to hydrodynamic description and to investigate the effects of the
different kinetic descriptions. The hybrid model is also used with a Chiral
EoS ($\chi$-EoS)~\cite{arXiv:0909.4421} featuring a rapid cross-over to
chirally restored and deconfined matter, which is compared to the
aforementioned hadronic descriptions and calculations with a MIT-Bag Model
EoS (BM-EoS) with a strong first-order phase transition. Both $\chi$-EoS and
BM-EoS have their phase transition at vanishing chemical potential at around
$T_{\sf C} \approx 170$~MeV.

\begin{table}
 \begin{tabular}{l|r}
  EoS & $\epsilon_{\sf crit}$ \\ \hline
  HG-EoS  & $5 \epsilon_0$ \\
  $\chi$-EoS & $7 \epsilon_0$ \\
  BM-EoS & $5 \epsilon_0$
 \end{tabular}
 \caption{The critical energy densities for the mapping from hydrodynamics
 to transport theory for the various Equations of state. $\epsilon_0 =
 146$~MeV/fm$^3$ is the nuclear ground state energy density.}
 \label{tab:epscrit}
\end{table}

\subsection{Intermediate stage in cascade calculations}

From the considerations above, it is clear that the hybrid model consists of
three stages, the pre-equilibrium early stage, the hydrodynamic intermediate
stage and the dilute late stage. When comparing the stages between the
different variations of the model and among each other, we assign the same
division of stages to the pure cascade calculations. Along these lines we
define the early stage as $0 < t < 3.7$~fm, the intermediate stage as $3.7 <
t < 11$~fm and the late stage after $t = 11$~fm.

\subsection{Photon emission sources}\label{sec:photons}

In the present work, direct photon emission is treated as a perturbation on
top of the evolution of the fireball, so the underlying evolution of the
hadronic medium remains unaltered by the calculation of direct photon
spectra. This is justified, because the ratio of electromagnetic
cross-sections (producing the photons) and strong cross-sections (governing
the evolution of the bulk) is very small.

The most important hadronic channels for the production of direct photons
are $\pi\pi\rightarrow\gamma\rho$ and $\pi\rho\rightarrow\gamma\pi$, which
both are implemented in the transport, as well as in the hydrodynamic phase.
The cross-sections for cascade-calculations are taken from Kapusta {\it et
al.}~\cite{Kapusta:1991qp}, while the rates used for the hydrodynamic
description have been parametrized by Turbide {\it et
al.}~\cite{Turbide:2003si}.  Since no thermal partonic interactions are
modelled in UrQMD, emission from a QGP-medium is only taken into account in
the hydrodynamic part of the model.  Several minor hadronic channels are
only implemented in one of the two models, such as strange channels (e.g.\
$K\pi\rightarrow\gamma{}K^\ast$) which are only present in the hydrodynamic
calculations, and $\eta$-channels (e.g.\ $\pi\eta\rightarrow\gamma\pi$)
which are only present in the transport calculations. Earlier investigations
with this model have shown those channels to contribute about equally, but
not significantly, to the overall spectrum of direct photons. The complete
list of channels and a detailed explanation of the calculation is provided
in~\cite{arXiv:0905.4678}.

\section{Results}\label{sec:results}

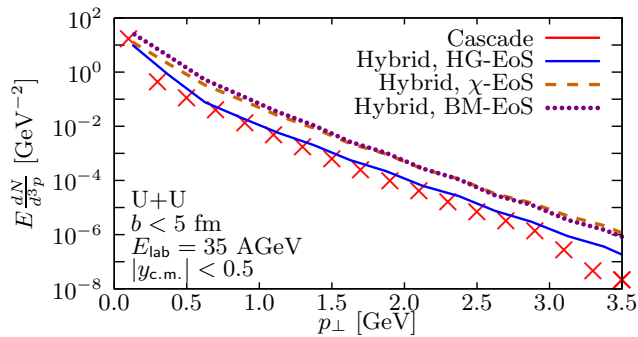
\begin{figure}
 \input{eoscomp}
 \caption{(Color Online) Overall direct photon spectra from
 U+U-collisions. Calculations in pure cascade mode are shown as red crosses,
 hybrid calculations with HG-EoS (see text) are shown as solid blue line,
 $\chi$-EoS-hybrid calculations as orange dashed line and BM-EoS-hybrid
 calculations are shown as purple dotted line.}
 \label{fig:inclusive}
\end{figure}

We start with a comparison of the overall direct photon spectra calculated
with the various Equations of State and with the transport-only approach.
From Figure~\ref{fig:inclusive}, one can clearly see that the transport-only
(crosses) and HG-EoS (solid line) calculations give very similar results.
The BM-EoS and $\chi$-EoS calculations, on the other hand, yield
significantly higher spectra than both hadronic calculations. The reason for
this difference will be discussed below.

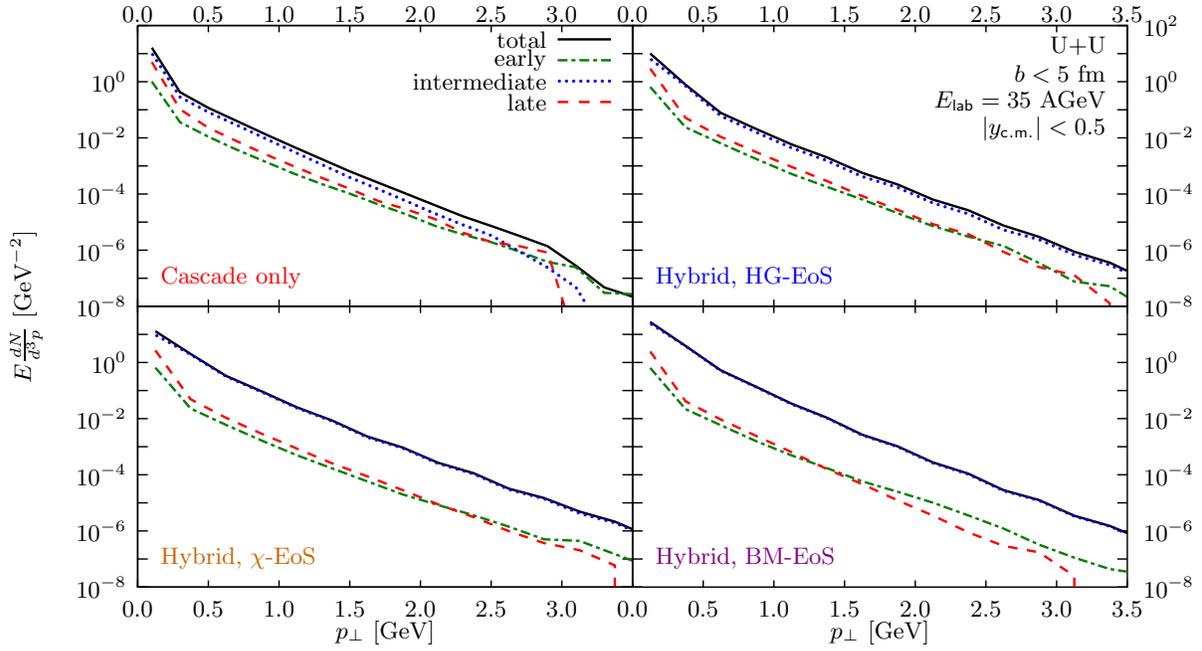
\begin{figure*}
 \input{stages}
 \caption{(Color Online) The contributions of the early (green dash-dotted
 lines), intermediate (blue dotted lines) and late (red dashed lines) stages to
 the overall direct photon spectra (solid black lines) separately for all four
 variations of the model. Calculations in transport mode are shown in the
 top left panel, hybrid calculations with HG-EoS on the top right,
 $\chi$-EoS hybrid calculations are in the bottom left, and BM-EoS hybrid
 calculations are in the bottom right panel.}
 \label{fig:stages}
\end{figure*}

Next, we investigate the contributions of the early, intermediate and late
stages to the overall spectra for all variations of the model. In
Figure~\ref{fig:stages}, we show that the intermediate stage, which is
calculated with hydrodynamics in the hybrid calculations, dominates the
emission in all cases, although the excess of this stage is less significant
in the hadronic calculations of pure transport and HG-EoS hybrid
calculations.

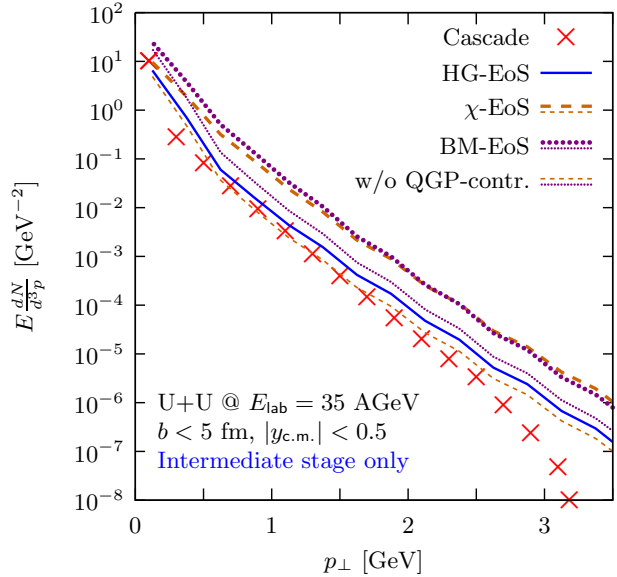
\begin{figure}
 \input{hydro}
 \caption{(Color Online) Comparison of the contributions of the intermediate
 stages for all variations of the model. In addition, the hadronic
 contributions to the intermediate-stage emissions in $\chi$-Eos and BM-EoS
 are shown in thin lines.}
 \label{fig:hydro}
\end{figure}

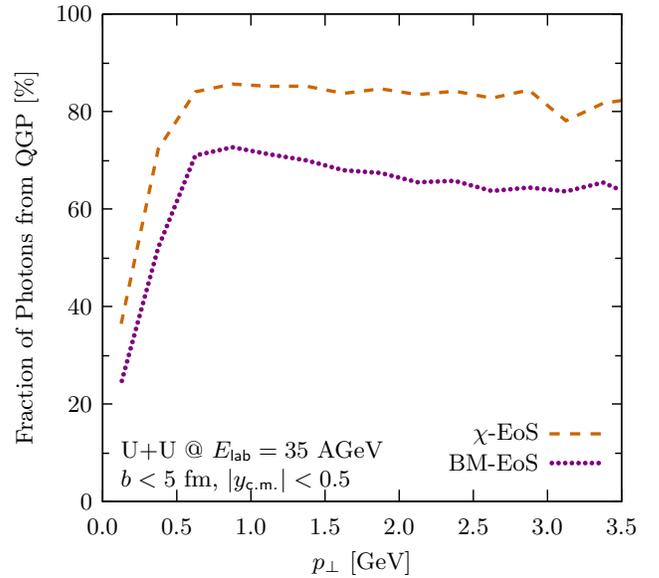
\begin{figure}
 \input{qgppart}
 \caption{(Color Online) The fraction of direct photons emitted from the
 QGP-phase in hybrid calculations with $\chi$-EoS (dashed orange line) and
 BM-EoS (purple dotted line) as a function of transverse photon momentum.}
 \label{fig:qgppart}
\end{figure}

Let us now take a look at the intermediate stages and investigate the
contribution of hadronic and partonic direct photon emissions to the spectra
from $\chi$-EoS and BM-EoS hybrid calculations. In Figure~\ref{fig:hydro},
we show the total contributions of the intermediate stages and the part of
it that comes from hadronic sources (in case of the $\chi$- and
BM-EoS-calculations). We note that the hadronic contributions are very
similar in all cases, and that the excess observed in calculations with
partonic degrees of freedom comes from the partonic part of the system.
Figure~\ref{fig:qgppart} shows the relative contribution of the QGP-phase in
hybrid calculations with BM- and $\chi$-EoS. The contribution remains fairly
constant at transverse momenta $p_\bot > 0.5$~GeV, and is at about $85~\%$
for the $\chi$-EoS and at about $65~\%$ for the BM-EoS-calculations.

\begin{table}
 \begin{tabular}{l|r@{.}l@{$\pm$}r@{.}l|r@{.}l@{$\pm$}r@{.}l|r@{.}l}
  Calculation & \multicolumn{4}{c|}{$T_{\sf slope}~[$MeV$]$} & \multicolumn{4}{c|}{$A~[$GeV$^{-2}]$} & \multicolumn{2}{c}{$\chi^2$/d.o.f.} \\ \hline
  Transport & 198&0 & 6&6 & 2&09 & 0&74 & 0&559 \\
  HG-EoS    & 203&5 & 8&0 & 2&98 & 1&16 & 0&532 \\
  $\chi$-EoS& 214&8 & 6&1 & 7&57 & 2&02 & 0&249 \\
  BM-EoS    & 200&8 & 5&8 &15&39 & 4&45 & 0&291 \\
 \end{tabular}
 \caption{Results of exponential fits to the spectra in the range $0 <
 p_\bot < 3.5$~GeV. The fit function is $f(p_\bot) = A \exp\left
 (-\frac{p_\bot}{T_{\sf slope}}\right )$. }
 \label{tab:temp}
\end{table}

What is eye-catching with the spectra is that despite the obvious
differences in magnitude, the slopes of the spectra are very similar.
Indeed, a closer anaylsis reveals all inverse slope parameters to be about
$T_{\sf slope} = 200$~MeV, with only the $\chi$-EoS calculations being
slightly higher (see Table~\ref{tab:temp}).

%

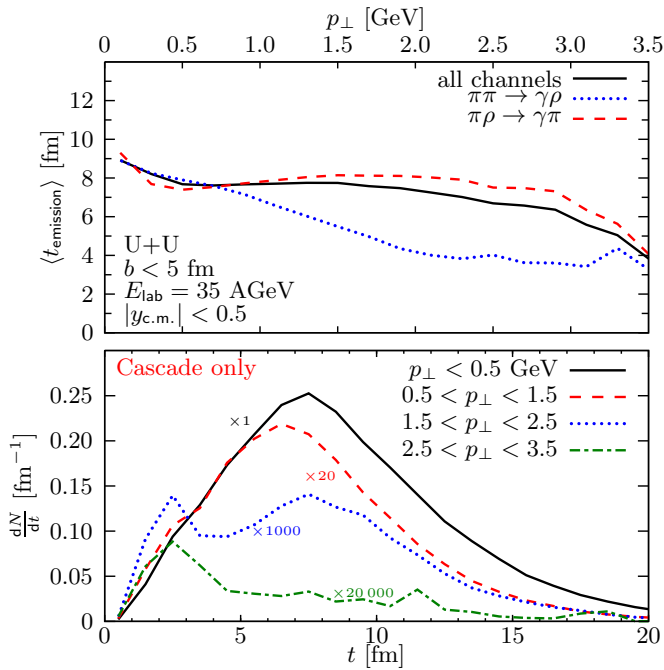
\begin{figure}
 \input{time}
 \caption{(Color Online) Average emission time of direct photons in cascade
 calculations as function of $p_\bot$ (top) and time evolution of the photon
 emission for various transverse momenta (bottom). The time evolution
 patterns from high transverse momenta have been scaled for readability.}
 \label{fig:time}
\end{figure}

\begin{figure}
 \input{timeparts}
 \caption{(Color Online) Fraction of photons that are emitted from $0 < t <
 5$~fm (red dashed line), $5 < t < 10$~fm (blue dotted line), $10 < t <
 15$~fm (green dash-dotted line), $15 < t < 20$~fm and $t > 20$~fm (violett
 dotted line) as a function of transverse momentum.}
 \label{fig:timeparts}
\end{figure}
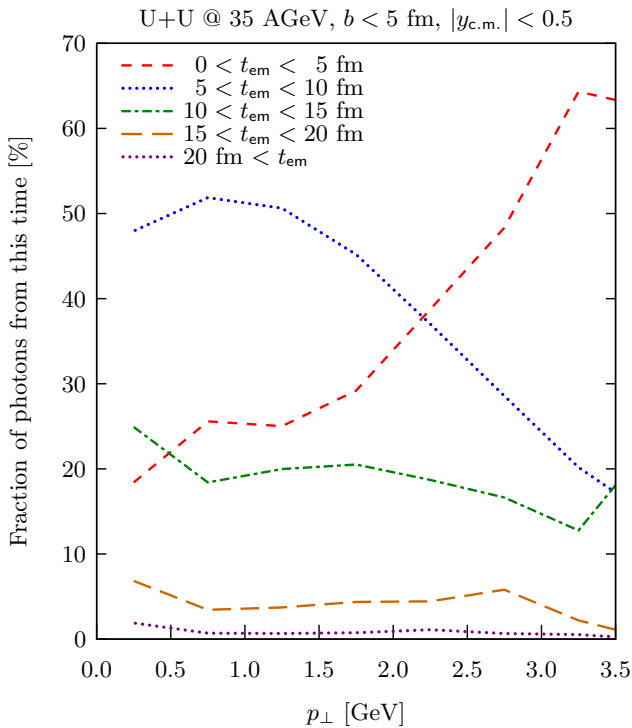

Since emission of direct photons happens throughout the heavy-ion collision,
it is prudent to look at the emission times of photons. We do so in
Figure~\ref{fig:time}, showing the average emission time of direct photons
plotted against their transverse momentum for the two most abundant channels
in cascade-only calculations and the time evolution of direct photon
emission for various transverse momentum bins.

We can see that the average emission time is nearly constant at about
$\langle t \rangle \approx 7.6$~fm over the whole $p_\bot$-range. It is
interesting to note that at intermediate $p_\bot$ direct photons from
$\pi\pi$-scatterings have a significantly lower average emission time
$\langle t_{\pi\pi} \rangle \approx 4$~fm than those coming from
$\pi\rho$-collisions.

The time evolution of direct photon emission in the lower part of
Figure~\ref{fig:time} shows that at intermediate transverse momentum $1.5 <
p_\bot < 2.5$~GeV the emission has two peaks, one after roughly $3$~fm and
the other one at $8$~fm. The average emission time in this momentum region
is $\langle t (p_\bot \approx 2\mbox{~GeV}) \rangle = 7.6$~fm. At high transverse
momentum, $2.5 < p_\bot < 3.5$~GeV, no significant peak at $t > 5$~fm is
present, but the tail of the distribution is still large enough to have the
average emission time as high as $\langle t (p_\bot \approx 3\mbox{~GeV}) \rangle =
6.4$~fm.

At very low $p_\bot < 0.5$~GeV, the emission is dominated by
$\pi\pi$-scatterings, and is symmetrically centered around its average value
of $\langle t (p_\bot \approx 0\mbox{~GeV}) \rangle = 8.8$~fm, as is the
$\pi\rho$-dominated photon emission at low transverse momentum $0.5 < p_\bot
< 1.5$~GeV, whose average time is $\langle t (p_\bot \approx 1\mbox{~GeV})
\rangle = 7.6$~fm.

When looking at the relative contributions of the different emission times
$t_{\sf em}$ to the direct photon yield as a function of direct photon
transverse momentum in Figure~\ref{fig:timeparts}, we can assert the picture
presented above. At low transverse momenta $p_\bot < 2$~GeV, emission from
$5 < t_{\sf em} < 10$~fm dominates and constitutes about 50~\% of the overall
contribution. At higher $p_\bot > 2.5$~GeV, however, emission from the very
early stages of the collision starts to dominate. The contribution of
photons emitted earlier than $t_{\sf em} = 5$~fm contribute more than 60~\%
to the direct photon spectrum at $p_\bot > 3$~GeV.

\begin{figure}
 \input{dndrho}
 \caption{Distribution of direct photons against the baryon number density
 $\rho_{\sf B}$ at the emission point for cascade calculations for all
 photons (red solid line) and for photons with $p_\bot > 2.5$~GeV (blue
 dotted line, scaled by $5 \cdot 10^4$).}
 \label{fig:dndrho}
\end{figure}
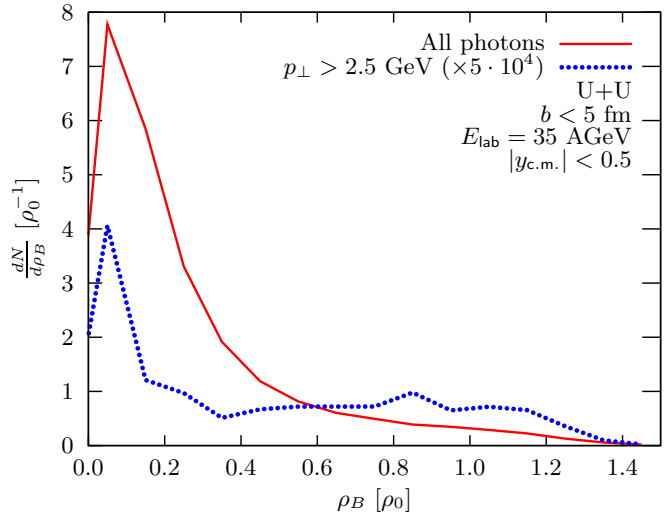

The baryon number density at the point of emission of direct photons is
investigated in Figure~\ref{fig:dndrho}. Most photons come from collisions
below $\rho_B < 0.5\rho_0$, with $\rho_0 = 0.16$~fm$^{-3}$ being the nuclear
ground state baryon number density. At high transverse momenta, the photon
emission is shifted from the low-density region towards high densities.  We
therefore assume it to be safe to employ vacuum cross-sections and vacuum
thermal rates which do not take into account high-density effects on the
$\rho$-spectral function.

\section{Summary}\label{sec:summary}

Direct photon emission from U+U-collisions at $E_{\sf lab} = 35$~AGeV with
$b < 5$~fm has been calculated and analysed within a microscopic transport
model and in a micro+macro hybrid model, in which the high-density phase of
the transport model has been replaced by ideal 3+1-dimensional hydrodynamic
calculations. In the hybrid framework, three different Equations of State
governing the behaviour of the matter in the hydrodynamic model have been
compared.

We find that the presence of partonic degrees of freedom significantly
enhance the amount of direct photons emitted from the medium, but not so
much the slope of the direct photon spectra. At the energy under
investigation, we find that this enhancement comes directly from the
partonic phase and is not due to a prolonged hadronic lifetime or hadronic
contributions from the mixed phase, which is present in the BM-EoS hybrid
calculations.

The average emission time of direct photons is nearly constant at $\langle t
\rangle \approx 7.6$~fm over a broad range of transverse momenta in the case
of cascade-only calculations, and also the spectra from hybrid model
calculations are dominated by emission from the intermediate stages. The
average emission time of photons originating from $\pi\pi$-scatterings drop
to lower times at transverse momenta larger than $p_\bot > 1.5$~GeV, but
since at this $p_\bot$-range this process is subleading to the dominant
$\pi\rho$-channel, this does not affect the overall average emission time.
Emission is found to come dominantly from areas with low baryon number
density. However, a trigger on high $p_\bot$ will allow to select photons
from the early stage.

\section{Outlook}

The predictions for FAIR-energies shown in this work lay the foundations for
future work analysing the ratio between direct photon emission and decay
photons. Further work with this model will include the calculation of direct
photon spectra at RHIC-energy and an analysis of the effect of changing the
interface-parameters between the transport and hydrodynamic models on the
direct photon emission.

\section{Acknowledgements}

This work has been supported by the Frankfurt Center for Scientific
Computing (CSC), the GSI and the BMBF. B.\ B\"auchle gratefully acknowledges
support from the Deutsche Telekom Stiftung, the Helmholtz Research School on
Quark Matter Studies and the Helmholtz Graduate School for Hadron and Ion
Research. This work was supported by the Hessian LOEWE initiative through
the Helmholtz International Center for FAIR.

The authors thank Elvira Santini for valuable discussions as well as Lars
Zeidlewicz and Tim Schuster.


\end{document}

%% file: eoscomp.tex
\ifx\PSTloaded\undefined
\def\PSTloaded{t}
\psset{arrowsize=.01 3.2 1.4 .3}
\psset{dotsize=.01}
\catcode`@=11

\newpsobject{PST@Border}{psline}{linewidth=.0015,linestyle=solid}
\newpsobject{PST@Axes}{psline}{linewidth=.0015,linestyle=dotted,dotsep=.004}
\newpsobject{PST@Solid}{psline}{linewidth=.0015,linestyle=solid}
\newpsobject{PST@Dashed}{psline}{linewidth=.0015,linestyle=dashed,dash=.01 .01}
\newpsobject{PST@Dotted}{psline}{linewidth=.0025,linestyle=dotted,dotsep=.008}
\newpsobject{PST@LongDash}{psline}{linewidth=.0015,linestyle=dashed,dash=.02 .01}
\newpsobject{PST@Diamond}{psdots}{linewidth=.001,linestyle=solid,dotstyle=square,dotangle=45}
\newpsobject{PST@Filldiamond}{psdots}{linewidth=.001,linestyle=solid,dotstyle=square*,dotangle=45}
\newpsobject{PST@Cross}{psdots}{linewidth=.001,linestyle=solid,dotstyle=+,dotangle=45}
\newpsobject{PST@Plus}{psdots}{linewidth=.001,linestyle=solid,dotstyle=+}
\newpsobject{PST@Square}{psdots}{linewidth=.001,linestyle=solid,dotstyle=square}
\newpsobject{PST@Circle}{psdots}{linewidth=.001,linestyle=solid,dotstyle=o}
\newpsobject{PST@Triangle}{psdots}{linewidth=.001,linestyle=solid,dotstyle=triangle}
\newpsobject{PST@Pentagon}{psdots}{linewidth=.001,linestyle=solid,dotstyle=pentagon}
\newpsobject{PST@Fillsquare}{psdots}{linewidth=.001,linestyle=solid,dotstyle=square*}
\newpsobject{PST@Fillcircle}{psdots}{linewidth=.001,linestyle=solid,dotstyle=*}
\newpsobject{PST@Filltriangle}{psdots}{linewidth=.001,linestyle=solid,dotstyle=triangle*}
\newpsobject{PST@Fillpentagon}{psdots}{linewidth=.001,linestyle=solid,dotstyle=pentagon*}
\newpsobject{PST@Arrow}{psline}{linewidth=.001,linestyle=solid}
\catcode`@=12

\fi
\psset{unit=5.0in,xunit=1.0638\columnwidth,yunit=.5\columnwidth}
\pspicture(0.0600,-.0100)(1.0000,1.0100)
\ifx\nofigs\undefined
\catcode`@=11

\multido{\i=22600+10486,\n=0.0+0.5}{8}{
\PST@Border(0.\i,0.1344) (0.\i,0.1544)
\PST@Border(0.\i,0.9680) (0.\i,0.9480)
\rput(0.\i,0.0920){\n}
}
\multido{\i=13440+8336}{11}{
 \PST@Border(0.2260,0.\i)(0.2410,0.\i)
 \PST@Border(0.9600,0.\i)(0.9450,0.\i)
}
\multido{\i=13440+16672,\ix=-8+2}{6}{
 \rput[r](0.2100,0.\i){$10^{\ix}$}
}

\PST@Border(0.2260,0.9680) (0.2260,0.1344) (0.9600,0.1344) (0.9600,0.9680)
(0.2260,0.9680)

\rput{L}(0.1010,0.5512){$E \frac{dN}{d^3p}$~[GeV$^{-2}$]}
\rput(0.5930,0.0294){$p_\bot$~[GeV]}

\rput[r](0.8330,0.9070){Cascade}
\rput[r](0.8330,0.8350){Hybrid, HG-EoS}
\rput[r](0.8330,0.7630){Hybrid, $\chi$-EoS}
\rput[r](0.8330,0.6910){Hybrid, BM-EoS}
\PST@cascl(0.8490,0.9070)(0.9280,0.9070)
\PST@hgeos(0.8490,0.8350)(0.9280,0.8350)
\PST@xieos(0.8490,0.7630)(0.9280,0.7630)
\PST@bmeos(0.8490,0.6910)(0.9280,0.6910)

\PST@cascd(0.2470,0.9023) (0.2889,0.7700) (0.3309,0.7235)
(0.3728,0.6848) (0.4147,0.6460) (0.4567,0.6084) (0.4986,0.5718) (0.5406,0.5351)
(0.5825,0.5017) (0.6245,0.4685) (0.6664,0.4349) (0.7083,0.4017) (0.7503,0.3731)
(0.7922,0.3435) (0.8342,0.3128) (0.8761,0.2548) (0.9181,0.1905) (0.9600,0.1632)
(0.9600,0.1632)

\PST@hgeos(0.2522,0.8851) (0.3046,0.7926) (0.3571,0.7086) (0.4095,0.6609)
(0.4619,0.6152) (0.5144,0.5766) (0.5668,0.5306) (0.6192,0.4962) (0.6716,0.4514)
(0.7241,0.4196) (0.7765,0.3741) (0.8289,0.3408) (0.8814,0.2965) (0.9338,0.2642)
(0.9600,0.2395)

\PST@xieos(0.2522,0.8941) (0.3046,0.8264) (0.3571,0.7615) (0.4095,0.7159)
(0.4619,0.6684) (0.5144,0.6290) (0.5668,0.5814) (0.6192,0.5487) (0.6716,0.5036)
(0.7241,0.4727) (0.7765,0.4267) (0.8289,0.3987) (0.8814,0.3586) (0.9338,0.3287)
(0.9600,0.3061)

\PST@bmeos(0.2522,0.9216) (0.3046,0.8505) (0.3571,0.7777)
(0.4095,0.7291) (0.4619,0.6791) (0.5144,0.6373) (0.5668,0.5867) (0.6192,0.5523)
(0.6716,0.5046) (0.7241,0.4712) (0.7765,0.4221) (0.8289,0.3929) (0.8814,0.3462)
(0.9338,0.3161) (0.9600,0.2945)

\rput[l](0.2500,0.4060){U+U}
\rput[l](0.2500,0.3340){$b < 5$ fm}
\rput[l](0.2500,0.2620){$E_{\sf lab} = 35$~AGeV}
\rput[l](0.2500,0.1900){$|y_{\sf c.m.}| < 0.5$}

\PST@Border(0.2260,0.9680) (0.2260,0.1344) (0.9600,0.1344) (0.9600,0.9680)
(0.2260,0.9680)

\catcode`@=12
\fi
\endpspicture

%% file: stages.tex
\ifx\PSTloaded\undefined
\def\PSTloaded{t}
\psset{arrowsize=.01 3.2 1.4 .3}
\psset{dotsize=.01}
\catcode`@=11

\newpsobject{PST@Border}{psline}{linewidth=.0015,linestyle=solid}
\newpsobject{PST@Axes}{psline}{linewidth=.0015,linestyle=dotted,dotsep=.004}
\newpsobject{PST@Solid}{psline}{linewidth=.0015,linestyle=solid}
\newpsobject{PST@Dashed}{psline}{linewidth=.0015,linestyle=dashed,dash=.01 .01}
\newpsobject{PST@Dotted}{psline}{linewidth=.0025,linestyle=dotted,dotsep=.008}
\newpsobject{PST@LongDash}{psline}{linewidth=.0015,linestyle=dashed,dash=.02 .01}
\newpsobject{PST@Diamond}{psdots}{linewidth=.001,linestyle=solid,dotstyle=square,dotangle=45}
\newpsobject{PST@Filldiamond}{psdots}{linewidth=.001,linestyle=solid,dotstyle=square*,dotangle=45}
\newpsobject{PST@Cross}{psdots}{linewidth=.001,linestyle=solid,dotstyle=+,dotangle=45}
\newpsobject{PST@Plus}{psdots}{linewidth=.001,linestyle=solid,dotstyle=+}
\newpsobject{PST@Square}{psdots}{linewidth=.001,linestyle=solid,dotstyle=square}
\newpsobject{PST@Circle}{psdots}{linewidth=.001,linestyle=solid,dotstyle=o}
\newpsobject{PST@Triangle}{psdots}{linewidth=.001,linestyle=solid,dotstyle=triangle}
\newpsobject{PST@Pentagon}{psdots}{linewidth=.001,linestyle=solid,dotstyle=pentagon}
\newpsobject{PST@Fillsquare}{psdots}{linewidth=.001,linestyle=solid,dotstyle=square*}
\newpsobject{PST@Fillcircle}{psdots}{linewidth=.001,linestyle=solid,dotstyle=*}
\newpsobject{PST@Filltriangle}{psdots}{linewidth=.001,linestyle=solid,dotstyle=triangle*}
\newpsobject{PST@Fillpentagon}{psdots}{linewidth=.001,linestyle=solid,dotstyle=pentagon*}
\newpsobject{PST@Arrow}{psline}{linewidth=.001,linestyle=solid}
\catcode`@=12

\fi
\psset{unit=5.0in,xunit=.50\textwidth,yunit=.25\textwidth}
\pspicture(0.0000,-.8836)(1.8340,1.0700)
\ifx\nofigs\undefined
\catcode`@=11

\multido{\i=22600+10486,\n=0.0+0.5}{7}{
\PST@Border(0.\i,0.1344) (0.\i,0.1544)
\PST@Border(0.\i,0.9680) (0.\i,0.9480)
\rput(0.\i,1.0000){\small \n}
}
\multido{\i=13440+8336}{11}{
 \PST@Border(0.2260,0.\i)(0.2410,0.\i)
 \PST@Border(0.9600,0.\i)(0.9450,0.\i)
}
\multido{\i=13440+16672,\ix=-8+2}{5}{
 \rput[r](0.2100,0.\i){\small $10^{\ix}$}
}

\PST@Border(0.2260,0.9680)
(0.2260,0.1344)
(0.9600,0.1344)
(0.9600,0.9680)
(0.2260,0.9680)

\rput{L}(0.0610,0.1344){\small $E \frac{dN}{d^3p}$~[GeV$^{-2}$]}

\PST@inc(0.8490,0.9270) (0.9280,0.9270)
\PST@bef(0.8490,0.8650) (0.9280,0.8650)
\PST@hyd(0.8490,0.8030) (0.9280,0.8030)
\PST@aft(0.8490,0.7410) (0.9280,0.7410)
\rput[r](0.8330,0.9270){\small total}
\rput[r](0.8330,0.8650){\small early}
\rput[r](0.8330,0.8030){\small intermediate}
\rput[r](0.8330,0.7410){\small late}
\rput[l](0.2600,0.2200){\small \color{red} Cascade only}

\PST@inc(0.2470,0.9023) (0.2470,0.9023) (0.2889,0.7700) (0.3309,0.7235)
(0.3728,0.6848) (0.4147,0.6460) (0.4567,0.6084) (0.4986,0.5718) (0.5406,0.5351)
(0.5825,0.5017) (0.6245,0.4685) (0.6664,0.4349) (0.7083,0.4017) (0.7503,0.3731)
(0.7922,0.3435) (0.8342,0.3128) (0.8761,0.2548) (0.9181,0.1905) (0.9600,0.1632)
(0.9600,0.1632)

\PST@bef(0.2470,0.8017) (0.2470,0.8017) (0.2889,0.6808) (0.3309,0.6379)
(0.3728,0.5994) (0.4147,0.5643) (0.4567,0.5292) (0.4986,0.4977) (0.5406,0.4690)
(0.5825,0.4374) (0.6245,0.4075) (0.6664,0.3737) (0.7083,0.3478) (0.7503,0.3249)
(0.7922,0.2971) (0.8342,0.2662) (0.8761,0.2503) (0.9181,0.1743) (0.9600,0.1706)
(0.9600,0.1706)

\PST@hyd(0.2470,0.8861) (0.2470,0.8861) (0.2889,0.7557) (0.3309,0.7108)
(0.3728,0.6725) (0.4147,0.6333) (0.4567,0.5943) (0.4986,0.5559) (0.5406,0.5171)
(0.5825,0.4826) (0.6245,0.4468) (0.6664,0.4103) (0.7083,0.3764) (0.7503,0.3449)
(0.7922,0.2972) (0.8342,0.2493) (0.8761,0.1909) (0.8932,0.1344)

\PST@aft(0.2470,0.8595) (0.2470,0.8595) (0.2889,0.7192) (0.3309,0.6644)
(0.3728,0.6249) (0.4147,0.5865) (0.4567,0.5508) (0.4986,0.5163) (0.5406,0.4832)
(0.5825,0.4475) (0.6245,0.4224) (0.6664,0.3939) (0.7083,0.3533) (0.7503,0.3253)
(0.7922,0.3130) (0.8342,0.2939) (0.8592,0.1344)

\PST@Border(0.2260,0.9680)
(0.2260,0.1344)
(0.9600,0.1344)
(0.9600,0.9680)
(0.2260,0.9680)

\catcode`@=12
\fi
\ifx\nofigs\undefined
\catcode`@=11

\PST@Border(0.9600,0.1344) (0.9600,0.1544)
\PST@Border(0.9600,0.9680) (0.9600,0.9480)
\rput(0.9600,1.0000){\small 0.0}
\PST@Border(1.0649,0.1344) (1.0649,0.1544)
\PST@Border(1.0649,0.9680) (1.0649,0.9480)
\rput(1.0649,1.0000){\small 0.5}
\multido{\i=16972+10486,\n=1.0+0.5}{6}{
\PST@Border(1.\i,0.1344) (1.\i,0.1544)
\PST@Border(1.\i,0.9680) (1.\i,0.9480)
\rput(1.\i,1.0000){\small \n}
}
\multido{\i=13440+8336}{11}{
 \PST@Border(0.9600,0.\i)(0.9750,0.\i)
 \PST@Border(1.6940,0.\i)(1.6790,0.\i)
}
\multido{\i=13440+16672,\ix=-8+2}{6}{
 \rput[l](1.7180,0.\i){\small $10^{\ix}$}
}

\PST@Border(0.9600,0.9680)
(0.9600,0.1344)
(1.6940,0.1344)
(1.6940,0.9680)
(0.9600,0.9680)

\rput[r](1.6620,0.9070){\small U+U}
\rput[r](1.6620,0.8250){\small $b < 5$~fm}
\rput[r](1.6620,0.7430){\small $E_{\sf lab} = 35$~AGeV}
\rput[r](1.6620,0.6610){\small $|y_{\sf c.m.}| < 0.5$}
\rput[l](0.9940,0.2200){\small \color{blue} Hybrid, HG-EoS}

\PST@inc(0.9862,0.8851) (1.0386,0.7926) (1.0911,0.7086) (1.1435,0.6609)
(1.1959,0.6152) (1.2484,0.5766) (1.3008,0.5306) (1.3532,0.4962) (1.4056,0.4514)
(1.4581,0.4196) (1.5105,0.3741) (1.5629,0.3408) (1.6154,0.2965) (1.6678,0.2642)
(1.6940,0.2395)

\PST@bef(0.9862,0.7858) (1.0386,0.6659) (1.0911,0.6181) (1.1435,0.5719)
(1.1959,0.5274) (1.2484,0.4891) (1.3008,0.4516) (1.3532,0.4109) (1.4056,0.3726)
(1.4581,0.3414) (1.5105,0.3133) (1.5629,0.2631) (1.6154,0.2079) (1.6678,0.1941)
(1.6940,0.1622)

\PST@hyd(0.9862,0.8693) (1.0386,0.7889) (1.0911,0.6995) (1.1435,0.6512)
(1.1959,0.6049) (1.2484,0.5670) (1.3008,0.5196) (1.3532,0.4873) (1.4056,0.4411)
(1.4581,0.4089) (1.5105,0.3611) (1.5629,0.3327) (1.6154,0.2869) (1.6678,0.2570)
(1.6940,0.2340)

\PST@aft(0.9862,0.8402) (1.0386,0.6941) (1.0911,0.6375) (1.1435,0.5923)
(1.1959,0.5484) (1.2484,0.5063) (1.3008,0.4619) (1.3532,0.4210) (1.4056,0.3785)
(1.4581,0.3487) (1.5105,0.2959) (1.5629,0.2510) (1.6154,0.2267) (1.6678,0.1437)
(1.6678,0.1344)

\PST@Border(0.9600,0.9680)
(0.9600,0.1344)
(1.6940,0.1344)
(1.6940,0.9680)
(0.9600,0.9680)

\catcode`@=12
\fi

\ifx\nofigs\undefined
\catcode`@=11

\multido{\i=22600+10486,\n=0.0+0.5}{7}{
\PST@Border(0.\i,-0.6992) (0.\i,-0.6792)
\PST@Border(0.\i,0.1344) (0.\i,0.1144)
\rput(0.\i,-0.7636){\small \n}
}
\multido{\n=-0.69920+0.08336}{11}{
 \PST@Border(0.2260,\n)(0.2410,\n)
 \PST@Border(0.9600,\n)(0.9450,\n)
}
\multido{\n=-0.69920+0.16672,\ix=-8+2}{5}{
 \rput[r](0.2100,\n){\small $10^{\ix}$}
}

\PST@Border(0.2260,0.1344)
(0.2260,-0.6992)
(0.9600,-0.6992)
(0.9600,0.1344)
(0.2260,0.1344)

\rput(0.5933,-0.8336){\small $p_\bot$ [GeV]}
\rput[l](0.2600,-0.6136){\small \color{orange} Hybrid, $\chi$-EoS}

\PST@inc(0.2522,0.0605) (0.3046,-0.0072) (0.3571,-0.0721) (0.4095,-0.1177)
(0.4619,-0.1652) (0.5144,-0.2046) (0.5668,-0.2522) (0.6192,-0.2849) (0.6716,-0.3300)
(0.7241,-0.3609) (0.7765,-0.4069) (0.8289,-0.4349) (0.8814,-0.4750) (0.9338,-0.5049)
(0.9600,-0.5275)

\PST@bef(0.2522,-0.0476) (0.3046,-0.1690) (0.3571,-0.2177) (0.4095,-0.2638)
(0.4619,-0.3076) (0.5144,-0.3468) (0.5668,-0.3855) (0.6192,-0.4242) (0.6716,-0.4556)
(0.7241,-0.4858) (0.7765,-0.5198) (0.8289,-0.5574) (0.8814,-0.5621) (0.9338,-0.6013)
(0.9600,-0.6211)

\PST@hyd(0.2522,0.0497) (0.3046,-0.0086) (0.3571,-0.0739) (0.4095,-0.1195)
(0.4619,-0.1672) (0.5144,-0.2065) (0.5668,-0.2546) (0.6192,-0.2868) (0.6716,-0.3324)
(0.7241,-0.3631) (0.7765,-0.4097) (0.8289,-0.4371) (0.8814,-0.4802) (0.9338,-0.5086)
(0.9600,-0.5309)

\PST@aft(0.2522,0.0036) (0.3046,-0.1416) (0.3571,-0.1967) (0.4095,-0.2435)
(0.4619,-0.2878) (0.5144,-0.3317) (0.5668,-0.3705) (0.6192,-0.4114) (0.6716,-0.4535)
(0.7241,-0.4914) (0.7765,-0.5330) (0.8289,-0.5686) (0.8814,-0.5896) (0.9338,-0.6348)
(0.9338,-0.6992)

\PST@Border(0.2260,0.1344)
(0.2260,-0.6992)
(0.9600,-0.6992)
(0.9600,0.1344)
(0.2260,0.1344)
\catcode`@=12
\fi
\ifx\nofigs\undefined
\catcode`@=11

\multido{\N=0.96000+0.10486,\n=0.0+0.5}{8}{
\PST@Border(\N,-.6992) (\N,-.6792)
\PST@Border(\N,0.1344) (\N,0.1144)
\rput(\N,-.7636){\small \n}
}
\multido{\n=-0.69920+0.08336}{11}{
 \PST@Border(0.9600,\n)(0.9750,\n)
 \PST@Border(1.6940,\n)(1.6790,\n)
}
\multido{\n=-0.69920+0.16672,\ix=-8+2}{5}{
 \rput[l](1.7180,\n){\small $10^{\ix}$}
}

\PST@Border(0.9600,0.1344)
(0.9600,-0.6992)
(1.6940,-0.6992)
(1.6940,0.1344)
(0.9600,0.1344)

\rput(1.3273,-0.8336){\small $p_\bot$ [GeV]}
\rput[l](0.9940,-0.6136){\small \color{violett} Hybrid, BM-EoS}

\PST@inc(0.9862,0.0880) (1.0386,0.0169) (1.0911,-0.0559) (1.1435,-0.1045)
(1.1959,-0.1545) (1.2484,-0.1963) (1.3008,-0.2469) (1.3532,-0.2813) (1.4056,-0.3290)
(1.4581,-0.3624) (1.5105,-0.4115) (1.5629,-0.4407) (1.6154,-0.4874) (1.6678,-0.5175)
(1.6940,-0.5391)

\PST@bef(0.9862,-0.0484) (1.0386,-0.1708) (1.0911,-0.2188) (1.1435,-0.2660)
(1.1959,-0.3081) (1.2484,-0.3460) (1.3008,-0.3853) (1.3532,-0.4161) (1.4056,-0.4486)
(1.4581,-0.4864) (1.5105,-0.5256) (1.5629,-0.5732) (1.6154,-0.6119) (1.6678,-0.6456)
(1.6940,-0.6535)

\PST@hyd(0.9862,0.0837) (1.0386,0.0163) (1.0911,-0.0569) (1.1435,-0.1056)
(1.1959,-0.1557) (1.2484,-0.1974) (1.3008,-0.2483) (1.3532,-0.2826) (1.4056,-0.3308)
(1.4581,-0.3639) (1.5105,-0.4134) (1.5629,-0.4422) (1.6154,-0.4889) (1.6678,-0.5186)
(1.6940,-0.5408)

\PST@aft(0.9862,0.0005) (1.0386,-0.1473) (1.0911,-0.2018) (1.1435,-0.2504)
(1.1959,-0.2983) (1.2484,-0.3490) (1.3008,-0.3935) (1.3532,-0.4417) (1.4056,-0.4867)
(1.4581,-0.5366) (1.5105,-0.5771) (1.5629,-0.5959) (1.6154,-0.6665) (1.6154,-0.6992)

\catcode`@=12
\fi
\endpspicture

%% file: hydro.tex
\ifx\PSTloaded\undefined
\def\PSTloaded{t}
\psset{arrowsize=.01 3.2 1.4 .3}
\psset{dotsize=.01}
\catcode`@=11

\newpsobject{PST@Border}{psline}{linewidth=.0015,linestyle=solid}
\newpsobject{PST@Axes}{psline}{linewidth=.0015,linestyle=dotted,dotsep=.004}
\newpsobject{PST@Solid}{psline}{linewidth=.0015,linestyle=solid}
\newpsobject{PST@Dashed}{psline}{linewidth=.0015,linestyle=dashed,dash=.01 .01}
\newpsobject{PST@Dotted}{psline}{linewidth=.0025,linestyle=dotted,dotsep=.008}
\newpsobject{PST@LongDash}{psline}{linewidth=.0015,linestyle=dashed,dash=.02 .01}
\newpsobject{PST@Diamond}{psdots}{linewidth=.001,linestyle=solid,dotstyle=square,dotangle=45}
\newpsobject{PST@Filldiamond}{psdots}{linewidth=.001,linestyle=solid,dotstyle=square*,dotangle=45}
\newpsobject{PST@Cross}{psdots}{linewidth=.001,linestyle=solid,dotstyle=+,dotangle=45}
\newpsobject{PST@Plus}{psdots}{linewidth=.001,linestyle=solid,dotstyle=+}
\newpsobject{PST@Square}{psdots}{linewidth=.001,linestyle=solid,dotstyle=square}
\newpsobject{PST@Circle}{psdots}{linewidth=.001,linestyle=solid,dotstyle=o}
\newpsobject{PST@Triangle}{psdots}{linewidth=.001,linestyle=solid,dotstyle=triangle}
\newpsobject{PST@Pentagon}{psdots}{linewidth=.001,linestyle=solid,dotstyle=pentagon}
\newpsobject{PST@Fillsquare}{psdots}{linewidth=.001,linestyle=solid,dotstyle=square*}
\newpsobject{PST@Fillcircle}{psdots}{linewidth=.001,linestyle=solid,dotstyle=*}
\newpsobject{PST@Filltriangle}{psdots}{linewidth=.001,linestyle=solid,dotstyle=triangle*}
\newpsobject{PST@Fillpentagon}{psdots}{linewidth=.001,linestyle=solid,dotstyle=pentagon*}
\newpsobject{PST@Arrow}{psline}{linewidth=.001,linestyle=solid}
\catcode`@=12

\fi
\psset{unit=5.0in,xunit=\columnwidth,yunit=.9\columnwidth}
\pspicture(0.0000,-.0100)(0.9600,1.0100)
\ifx\nofigs\undefined
\catcode`@=11

\multido{\i=22600+10486}{8}{
\PST@Border(0.\i,0.1344) (0.\i,0.1544)
\PST@Border(0.\i,0.9680) (0.\i,0.9480)
}
\multido{\i=22600+20972,\ix=0+1}{4}{
\rput(0.\i,0.0920){\ix}
}
\multido{\i=13440+8336,\ix=-8+1}{11}{
 \PST@Border(0.2260,0.\i)(0.2410,0.\i)
 \PST@Border(0.9600,0.\i)(0.9450,0.\i)
 \rput[r](0.2100,0.\i){\small $10^{\ix}$}
}

\PST@Border(0.2260,0.9680) (0.2260,0.1344) (0.9600,0.1344) (0.9600,0.9680)
(0.2260,0.9680)

\rput{L}(0.0610,0.5512){\small $E \frac{dN}{d^3p}$~[GeV$^{-2}$]}
\rput(0.5930,0.0294){$p_\bot$~[GeV]}

\PST@cascd(0.8885,0.9270)
\PST@hgeos(0.8490,0.8650) (0.9280,0.8650)
\PST@xieos(0.8490,0.8080) (0.9280,0.8080)
\PST@bmeos(0.8490,0.7460) (0.9280,0.7460)
\rput[r](0.8330,0.9270){\small Cascade}
\rput[r](0.8330,0.8650){\small HG-EoS}
\rput[r](0.8330,0.8030){\small $\chi$-EoS}
\rput[r](0.8330,0.7410){\small BM-EoS}
\rput[r](0.8330,0.6790){\small w/o QGP-contr.}
\rput[l](0.2600,0.3000){\small U+U @ $E_{\sf lab}=35$~AGeV}
\rput[l](0.2600,0.2500){\small $b < 5$~fm, $|y_{\sf c.m.}| < 0.5$}
\rput[l](0.2600,0.2000){\small\color{blue} Intermediate stage only}

\PST@cascd(0.2470,0.8861) (0.2470,0.8861) (0.2889,0.7557) (0.3309,0.7108)
(0.3728,0.6725) (0.4147,0.6333) (0.4567,0.5943) (0.4986,0.5559) (0.5406,0.5171)
(0.5825,0.4826) (0.6245,0.4468) (0.6664,0.4103) (0.7083,0.3764) (0.7503,0.3449)
(0.7922,0.2972) (0.8342,0.2493) (0.8761,0.1909) (0.8932,0.1344)

\PST@hgeos(0.2522,0.8693) (0.3046,0.7889) (0.3571,0.6995) (0.4095,0.6512)
(0.4619,0.6049) (0.5144,0.5670) (0.5668,0.5196) (0.6192,0.4873) (0.6716,0.4411)
(0.7241,0.4089) (0.7765,0.3611) (0.8289,0.3327) (0.8814,0.2869) (0.9338,0.2570)
(0.9600,0.2340)

\PST@xieos(0.2522,0.8833) (0.3046,0.8250) (0.3571,0.7597) (0.4095,0.7141)
(0.4619,0.6664) (0.5144,0.6271) (0.5668,0.5790) (0.6192,0.5468) (0.6716,0.5012)
(0.7241,0.4705) (0.7765,0.4239) (0.8289,0.3965) (0.8814,0.3534) (0.9338,0.3250)
(0.9600,0.3027)

\PST@bmeos(0.2522,0.9173) (0.3046,0.8499) (0.3571,0.7767) (0.4095,0.7280)
(0.4619,0.6779) (0.5144,0.6362) (0.5668,0.5853) (0.6192,0.5510) (0.6716,0.5028)
(0.7241,0.4697) (0.7765,0.4202) (0.8289,0.3914) (0.8814,0.3447) (0.9338,0.3150)
(0.9600,0.2928)

{
\psset{unit=2.5in,xunit=\columnwidth,yunit=.9\columnwidth}

\PST@xieos(0.8490,0.7980) (0.9280,0.7980)
\PST@bmeos(0.8490,0.7360) (0.9280,0.7360)
\PST@xieos(0.8490,0.6840) (0.9280,0.6840)
\PST@bmeos(0.8490,0.6740) (0.9280,0.6740)

\PST@xieos(0.2522,0.8589) (0.3046,0.7749) (0.3571,0.6815) (0.4095,0.6307)
(0.4619,0.5832) (0.5144,0.5451) (0.5668,0.4979) (0.6192,0.4659) (0.6716,0.4207)
(0.7241,0.3891) (0.7765,0.3425) (0.8289,0.3144) (0.8814,0.2700) (0.9338,0.2398)
(0.9600,0.2181)

\PST@bmeos(0.2522,0.9057) (0.3046,0.8226) (0.3571,0.7291) (0.4095,0.6781)
(0.4619,0.6296) (0.5144,0.5899) (0.5668,0.5408) (0.6192,0.5075) (0.6716,0.4605)
(0.7241,0.4278) (0.7765,0.3797) (0.8289,0.3512) (0.8814,0.3054) (0.9338,0.2744)
(0.9600,0.2529)
}

\PST@Border(0.2260,0.9680) (0.2260,0.1344) (0.9600,0.1344) (0.9600,0.9680)
(0.2260,0.9680)

\catcode`@=12
\fi
\endpspicture

%% file: qgppart.tex
\ifx\PSTloaded\undefined
\def\PSTloaded{t}
\psset{arrowsize=.01 3.2 1.4 .3}
\psset{dotsize=.01}
\catcode`@=11

\newpsobject{PST@Border}{psline}{linewidth=.0015,linestyle=solid}
\catcode`@=12

\fi
\psset{unit=5.0in,xunit=\columnwidth,yunit=.9\columnwidth}
\pspicture(0.000000,0.000000)(1.000000,1.000000)
\ifx\nofigs\undefined
\catcode`@=11

\multido{\i=1620+1140,\n=0.0+0.5}{8}{
\PST@Border(0.\i,0.1344) (0.\i,0.1544)
\PST@Border(0.\i,0.9680) (0.\i,0.9480)
\rput(0.\i,0.0920){\n}
}
\multido{\i=21675+16670}{5}{
 \PST@Border(0.1620,0.\i)(0.1720,0.\i)
 \PST@Border(0.9600,0.\i)(0.9500,0.\i)
}
\multido{\i=1344+1667,\ix=0+20}{6}{
 \PST@Border(0.1620,0.\i)(0.1770,0.\i)
 \PST@Border(0.9600,0.\i)(0.9450,0.\i)
 \rput[r](0.1460,0.\i){\small \ix}
}

\PST@Border(0.1620,0.9679) (0.1620,0.1344) (0.9599,0.1344) (0.9599,0.9679)
(0.1620,0.9679)

\rput{L}(0.0450,0.5511){Fraction of Photons from QGP [\%]}
\rput(0.5609,0.0294){$p_\bot$~[GeV]}

\PST@xieos(0.8490,0.2500) (0.9280,0.2500)
\PST@bmeos(0.8490,0.2000) (0.9280,0.2000)
\rput[r](0.8330,0.2500){\small $\chi$-EoS}
\rput[r](0.8330,0.2000){\small BM-EoS}
\rput[l](0.1900,0.2200){\small U+U @ $E_{\sf lab}=35$~AGeV}
\rput[l](0.1900,0.1700){\small $b < 5$~fm, $|y_{\sf c.m.}| < 0.5$}

\PST@bmeos(0.1905,0.3366) (0.1905,0.3366) (0.2475,0.5684) (0.3045,0.7260)
(0.3615,0.7400) (0.4185,0.7279) (0.4755,0.7175) (0.5325,0.7007) (0.5894,0.6960)
(0.6464,0.6802) (0.7034,0.6825) (0.7604,0.6652) (0.8174,0.6711) (0.8744,0.6645)
(0.9314,0.6797) (0.9599,0.6659)

\PST@xieos(0.1905,0.4381) (0.1905,0.4381) (0.2475,0.7369) (0.3045,0.8348)
(0.3615,0.8480) (0.4185,0.8442) (0.4755,0.8444) (0.5325,0.8321) (0.5894,0.8400)
(0.6464,0.8300) (0.7034,0.8359) (0.7604,0.8240) (0.8174,0.8379) (0.8744,0.7854)
(0.9314,0.8151) (0.9599,0.8199)

\PST@Border(0.1620,0.9679)
(0.1620,0.1344)
(0.9599,0.1344)
(0.9599,0.9679)
(0.1620,0.9679)

\catcode`@=12
\fi
\endpspicture

%% file: time.tex
\ifx\PSTloaded\undefined
\def\PSTloaded{t}
\psset{arrowsize=.01 3.2 1.4 .3}
\psset{dotsize=.01}
\catcode`@=11

\newpsobject{PST@Border}{psline}{linewidth=.0015,linestyle=solid}
\newpsobject{PST@Axes}{psline}{linewidth=.0015,linestyle=dotted,dotsep=.004}
\newpsobject{PST@Solid}{psline}{linewidth=.0015,linestyle=solid}
\newpsobject{PST@Dashed}{psline}{linewidth=.0015,linestyle=dashed,dash=.01 .01}
\newpsobject{PST@Dotted}{psline}{linewidth=.0025,linestyle=dotted,dotsep=.008}
\newpsobject{PST@LongDash}{psline}{linewidth=.0015,linestyle=dashed,dash=.02 .01}
\newpsobject{PST@Diamond}{psdots}{linewidth=.001,linestyle=solid,dotstyle=square,dotangle=45}
\newpsobject{PST@Filldiamond}{psdots}{linewidth=.001,linestyle=solid,dotstyle=square*,dotangle=45}
\newpsobject{PST@Cross}{psdots}{linewidth=.001,linestyle=solid,dotstyle=+,dotangle=45}
\newpsobject{PST@Plus}{psdots}{linewidth=.001,linestyle=solid,dotstyle=+}
\newpsobject{PST@Square}{psdots}{linewidth=.001,linestyle=solid,dotstyle=square}
\newpsobject{PST@Circle}{psdots}{linewidth=.001,linestyle=solid,dotstyle=o}
\newpsobject{PST@Triangle}{psdots}{linewidth=.001,linestyle=solid,dotstyle=triangle}
\newpsobject{PST@Pentagon}{psdots}{linewidth=.001,linestyle=solid,dotstyle=pentagon}
\newpsobject{PST@Fillsquare}{psdots}{linewidth=.001,linestyle=solid,dotstyle=square*}
\newpsobject{PST@Fillcircle}{psdots}{linewidth=.001,linestyle=solid,dotstyle=*}
\newpsobject{PST@Filltriangle}{psdots}{linewidth=.001,linestyle=solid,dotstyle=triangle*}
\newpsobject{PST@Fillpentagon}{psdots}{linewidth=.001,linestyle=solid,dotstyle=pentagon*}
\newpsobject{PST@Arrow}{psline}{linewidth=.001,linestyle=solid}
\catcode`@=12

\fi
\psset{unit=5.0in,xunit=1.0917\columnwidth,yunit=.5\columnwidth}
\pspicture(0.0500,0.1000)(0.9800,1.1400)
\ifx\nofigs\undefined
\catcode`@=11

\multido{\i=1344+1191}{7}{  \PST@Border(0.1940,0.\i)(0.2084,0.\i)\PST@Border(0.9600,0.\i)(0.9456,0.\i)}
\multido{\i=19395+11910}{7}{\PST@Border(0.1940,0.\i)(0.2036,0.\i)\PST@Border(0.9600,0.\i)(0.9504,0.\i)}
\multido{\i=1940+1094}{7}{  \PST@Border(0.\i,0.1344)(0.\i,0.1544)\PST@Border(0.\i,0.9680)(0.\i,0.9480)}

\multido{\i=1344+1191,\ix=0+2}{7}{\rput[r](0.1786,0.\i){\ix}}
\multido{\i=1940+1094,\nx=0+0.5}{8}{\rput(0.\i,1.0224){\nx}}

\PST@Border(0.1940,0.9680) (0.1940,0.1344) (0.9600,0.1344) (0.9600,0.9680)
(0.1940,0.9680)

\rput{L}(0.1201,0.5512){$\langle t_{\sf emission}\rangle$~[fm]}
\rput(0.5770,1.0994){$p_\bot$~[GeV]}
\rput[r](0.8381,0.9170){all channels}
\rput[r](0.8381,0.8550){$\pi\pi\rightarrow\gamma \rho$}
\rput[r](0.8381,0.7930){$\pi\rho\rightarrow\gamma \pi$}
\PST@inklusiv(0.8535,0.9170)(0.9293,0.9170)
\PST@pipigrho(0.8535,0.8550)(0.9293,0.8550)
\PST@pirhogpi(0.8535,0.7930)(0.9293,0.7930)

\rput[l](0.2209,0.4060){U+U}
\rput[l](0.2209,0.3340){$b < 5$ fm}
\rput[l](0.2209,0.2620){$E_{\sf lab} = 35$~AGeV}
\rput[l](0.2209,0.1900){$|y_{\sf c.m.}| < 0.5$}

\PST@inklusiv(0.2159,0.6656) (0.2159,0.6656) (0.2597,0.6231) (0.3034,0.5916)
(0.3472,0.5878) (0.3910,0.5912) (0.4347,0.5933) (0.4785,0.5959) (0.5223,0.5956)
(0.5661,0.5857) (0.6098,0.5796) (0.6536,0.5661) (0.6974,0.5522) (0.7411,0.5328)
(0.7849,0.5259) (0.8287,0.5132) (0.8725,0.4669) (0.9162,0.4348) (0.9600,0.3628)
(0.9600,0.3628)

\PST@pipigrho(0.2159,0.6650) (0.2159,0.6650) (0.2597,0.6253) (0.3034,0.6049)
(0.3472,0.5861) (0.3910,0.5602) (0.4347,0.5257) (0.4785,0.4935) (0.5223,0.4614)
(0.5661,0.4305) (0.6098,0.3937) (0.6536,0.3732) (0.6974,0.3624) (0.7411,0.3737)
(0.7849,0.3501) (0.8287,0.3491) (0.8725,0.3380) (0.9162,0.3944) (0.9600,0.3298)
(0.9600,0.3298)

\PST@pirhogpi(0.2159,0.6883) (0.2159,0.6883) (0.2597,0.5921) (0.3034,0.5747)
(0.3472,0.5830) (0.3910,0.5945) (0.4347,0.6039) (0.4785,0.6134) (0.5223,0.6193)
(0.5661,0.6179) (0.6098,0.6172) (0.6536,0.6124) (0.6974,0.6055) (0.7411,0.5814)
(0.7849,0.5796) (0.8287,0.5698) (0.8725,0.5124) (0.9162,0.4697) (0.9600,0.3771)
(0.9600,0.3771)

\PST@Border(0.1940,0.9680) (0.1940,0.1344) (0.9600,0.1344) (0.9600,0.9680)
(0.1940,0.9680)

\catcode`@=12
\fi
\endpspicture

\pspicture(0.0500,-.0100)(0.9800,0.9800)
\ifx\nofigs\undefined
\catcode`@=11

\multido{\i=1344+1389,\nx=0+.05}{6}{
 \PST@Border(0.1940,0.\i)(0.2090,0.\i)\PST@Border(0.9600,0.\i)(0.9450,0.\i)
 \rput[r](0.1780,0.\i){\nx}
}
\multido{\i=2323+383}{19}{
 \PST@Border(0.\i,0.1344)(0.\i,0.1444)\PST@Border(0.\i,0.9680)(0.\i,0.9580)
}
\multido{\i=1940+1915,\ix=0+5}{5}{
 \PST@Border(0.\i,0.1344)(0.\i,0.1544)\PST@Border(0.\i,0.9680)(0.\i,0.9480)
 \rput(0.\i,0.0924){\ix}
}

\PST@Border(0.1940,0.9680) (0.1940,0.1344) (0.9600,0.1344) (0.9600,0.9680)
(0.1940,0.9680)

\rput{L}(0.0810,0.5512){$\frac{{\sf d}N}{{\sf d}t}$~[fm$^{-1}$]}
\rput(0.5770,0.0294){$t$~[fm]}

\rput[l](0.2100,0.9100){\color{red}Cascade only}

\rput[r](0.8330,0.9070){\small $p_\bot < 0.5$~GeV}
\PST@inklusiv(0.8490,0.9070)(0.9280,0.9070)
\rput[r](0.4000,0.7500){\tiny \color{black}$\times 1$}
\PST@inklusiv(0.2132,0.1404) (0.2515,0.2487) (0.2898,0.3953) (0.3281,0.4927)
(0.3664,0.6154) (0.4047,0.7096) (0.4430,0.8002) (0.4813,0.8361) (0.5196,0.7796)
(0.5579,0.6864) (0.5962,0.6076) (0.6345,0.5246) (0.6728,0.4422) (0.7111,0.3815)
(0.7494,0.3267) (0.7877,0.2765) (0.8260,0.2440) (0.8643,0.2155) (0.9026,0.1953)
(0.9409,0.1783) (0.9600,0.1720)

\rput[r](0.8330,0.8250){\small $0.5 < p_\bot < 1.5$}
\PST@pirhogpi(0.8490,0.8250)(0.9280,0.8250)
\rput[r](0.5200,0.5800){\tiny \color{red}$\times 20$}
\PST@pirhogpi (0.2132,0.1435) (0.2515,0.2951) (0.2898,0.4299) (0.3281,0.4829)
(0.3664,0.6226) (0.4047,0.6964) (0.4430,0.7422) (0.4813,0.7105) (0.5196,0.6311)
(0.5579,0.5312) (0.5962,0.4527) (0.6345,0.3727) (0.6728,0.3106) (0.7111,0.2590)
(0.7494,0.2255) (0.7877,0.1989) (0.8260,0.1817) (0.8643,0.1647) (0.9026,0.1558)
(0.9409,0.1483) (0.9600,0.1463)

\rput[r](0.8330,0.7430){\small $1.5 < p_\bot < 2.5$}
\PST@pipigrho(0.8490,0.7430)(0.9280,0.7430)
\rput[l](0.4000,0.4100){\tiny \color{blue}$\times 1000$}
\PST@pipigrho (0.2132,0.1480) (0.2515,0.3881) (0.2898,0.5218) (0.3281,0.3982)
(0.3664,0.3952) (0.4047,0.4315) (0.4430,0.4894) (0.4813,0.5258) (0.5196,0.4861)
(0.5579,0.4614) (0.5962,0.3913) (0.6345,0.3392) (0.6728,0.2816) (0.7111,0.2375)
(0.7494,0.2127) (0.7877,0.1936) (0.8260,0.1779) (0.8643,0.1676) (0.9026,0.1559)
(0.9409,0.1472) (0.9600,0.1452)

\rput[r](0.8330,0.6610){\small $2.5 < p_\bot < 3.5$}
\PST@etachann(0.8490,0.6610)(0.9280,0.6610)
\rput[c](0.5579,0.2200){\tiny \color{darkgreen}$\times 20\,000$}
\PST@etachann(0.2132,0.1496) (0.2515,0.3039) (0.2898,0.3799) (0.3281,0.3067)
(0.3664,0.2282) (0.4047,0.2195) (0.4430,0.2124) (0.4813,0.2261) (0.5196,0.1955)
(0.5579,0.2023) (0.5962,0.1813) (0.6345,0.2322) (0.6728,0.1706) (0.7111,0.1634)
(0.7494,0.1495) (0.7877,0.1450) (0.8260,0.1435) (0.8643,0.1584) (0.9026,0.1652)
(0.9409,0.1350) (0.9600,0.1350)

\PST@Border(0.1940,0.9680) (0.1940,0.1344) (0.9600,0.1344) (0.9600,0.9680)
(0.1940,0.9680)

\catcode`@=12
\fi
\endpspicture

%% file: timeparts.tex
\ifx\PSTloaded\undefined
\def\PSTloaded{t}
\psset{arrowsize=.01 3.2 1.4 .3}
\psset{dotsize=.01}
\catcode`@=11

\newpsobject{PST@Border}{psline}{linewidth=.0015,linestyle=solid}
\catcode`@=12

\fi
\psset{unit=5.0in,xunit=\columnwidth,yunit=1.1\columnwidth}
\pspicture(0.000000,0.000000)(1.000000,1.000000)
\ifx\nofigs\undefined
\catcode`@=11

\multido{\i=1620+1140,\n=0.0+0.5}{8}{
\PST@Border(0.\i,0.1344) (0.\i,0.1544)
\rput(0.\i,0.0920){\n}
}
\multido{\i=134400000+119071428,\ix=0+10}{8}{
 \PST@Border(0.1620,0.\i)(0.1770,0.\i)
 \PST@Border(0.9600,0.\i)(0.9450,0.\i)
 \rput[r](0.1460,0.\i){\small \ix}
}

\PST@Border(0.1620,0.9679) (0.1620,0.1344) (0.9599,0.1344) (0.9599,0.9679)
(0.1620,0.9679)

\rput{L}(0.0450,0.5511){Fraction of photons from this time $[\%]$}
\rput(0.5609,0.0294){$p_\bot$~[GeV]}
\rput[l](0.2950,0.9370){\small $\phantom{0}0 < t_{\sf em} < \phantom{0}5$~fm}
\rput[l](0.2950,0.9050){\small $\phantom{0}5 < t_{\sf em} < 10$~fm}
\rput[l](0.2950,0.8730){\small           $10 < t_{\sf em} < 15$~fm}
\rput[l](0.2950,0.8410){\small           $15 < t_{\sf em} < 20$~fm}
\rput[l](0.2950,0.8090){\small     $20$~fm~$ < t_{\sf em}$}
\rput[c](0.5609,1.0000){\small U+U @ 35~AGeV, $b < 5$~fm, $|y_{\sf c.m.}| < 0.5$}

\PST@pirhogpi(0.2000,0.9370) (0.2790,0.9370)
\PST@pipigrho(0.2000,0.9050) (0.2790,0.9050)
\PST@etachann(0.2000,0.8730) (0.2790,0.8730)
\PST@gammagam(0.2000,0.8410) (0.2790,0.8410)
\PST@quglupla(0.2000,0.8090) (0.2790,0.8090) 

\PST@pirhogpi(0.2190,0.3536) (0.3330,0.4390) (0.4470,0.4323) (0.5610,0.4820)
(0.6749,0.5956) (0.7889,0.7099) (0.9029,0.9003) (0.9599,0.8886)

\PST@pipigrho(0.2190,0.7053) (0.3330,0.7520) (0.4470,0.7372) (0.5610,0.6724)
(0.6749,0.5753) (0.7889,0.4746) (0.9029,0.3747) (0.9599,0.3393)

\PST@etachann(0.2190,0.4311) (0.3330,0.3536) (0.4470,0.3723) (0.5610,0.3786)
(0.6749,0.3571) (0.7889,0.3325) (0.9029,0.2864) (0.9599,0.3497)

\PST@gammagam(0.2190,0.2158) (0.3330,0.1754) (0.4470,0.1787) (0.5610,0.1864)
(0.6749,0.1872) (0.7889,0.2035) (0.9029,0.1606) (0.9599,0.1475)

\PST@quglupla(0.2190,0.1570) (0.3330,0.1427) (0.4470,0.1423) (0.5610,0.1433)
(0.6749,0.1475) (0.7889,0.1422) (0.9029,0.1407) (0.9599,0.1375)

\PST@Border(0.1620,0.9679) (0.1620,0.1344) (0.9599,0.1344) (0.9599,0.9679)
(0.1620,0.9679)

\catcode`@=12
\fi
\endpspicture

%% file: dndrho.tex
\ifx\PSTloaded\undefined
\def\PSTloaded{t}
\psset{arrowsize=.01 3.2 1.4 .3}
\psset{dotsize=.01}
\catcode`@=11

\newpsobject{PST@Border}{psline}{linewidth=.0015,linestyle=solid}
\newpsobject{PST@all}{psline}{linecolor=red, linewidth=.0025,linestyle=solid}
\newpsobject{PST@ptc}{psline}{linecolor=blue,linewidth=.0050,linestyle=dotted,dotsep=.002}

\catcode`@=12

\fi
\psset{unit=5.0in,xunit=1.4136\columnwidth,yunit=.8\columnwidth}
\pspicture(0.0450,0.0000)(0.7524,0.9680)
\ifx\nofigs\undefined
\catcode`@=11

\multido{\i=0+1,\ia=1344+1042}{9}{
 \PST@Border(0.1300,0.\ia) (0.1450,0.\ia)
 \PST@Border(0.7524,0.\ia) (0.7324,0.\ia)
 \rput[r](0.1140,0.\ia){\i}
}
\multido{\n=0.0+0.2,\ia=1300+830}{8}{
 \PST@Border(0.\ia,0.1344) (0.\ia,0.1544)
 \PST@Border(0.\ia,0.9680) (0.\ia,0.9480)
 \rput(0.\ia,0.0924){\n}
}

\PST@Border(0.1300,0.9680) (0.1300,0.1344) (0.7524,0.1344) (0.7524,0.9680) (0.1300,0.9680)

\rput{L}(0.0650,0.5511){$\frac{dN}{d\rho_B}$ [$\rho_0^{-1}$]}
\rput(0.4412,0.0294){$\rho_B$ [$\rho_0$]}

\rput[r](0.6262,0.9070){All photons}
\rput[r](0.6262,0.8620){$p_\bot > 2.5$~GeV $(\times 5 \cdot 10^4)$}
\rput[r](0.7212,0.8170){U+U}
\rput[r](0.7212,0.7720){$b < 5$ fm}
\rput[r](0.7212,0.7270){$E_{\sf lab} = 35$~AGeV}
\rput[r](0.7212,0.6820){$|y_{\sf c.m.}| < 0.5$}
\PST@all(0.6423,0.9070) (0.7212,0.9070)
\PST@ptc(0.6423,0.8620) (0.7212,0.8620)

\PST@all(0.1300,0.5396) (0.1507,0.9445) (0.1922,0.7436) (0.2337,0.4792) (0.2752,0.3340)
(0.3167,0.2583) (0.3582,0.2193) (0.3997,0.1973) (0.4412,0.1859) (0.4827,0.1749)
(0.5242,0.1705) (0.5657,0.1645) (0.6072,0.1578) (0.6487,0.1478) (0.6902,0.1404)
(0.7317,0.1364)

\PST@ptc(0.1300,0.3460) (0.1507,0.5576) (0.1922,0.2605) (0.2337,0.2354)
(0.2752,0.1877) (0.3167,0.2041) (0.3582,0.2094) (0.3997,0.2094) (0.4412,0.2095)
(0.4827,0.2363) (0.5242,0.2021) (0.5657,0.2091) (0.6072,0.2026) (0.6487,0.1716)
(0.6902,0.1446) (0.7317,0.1366)

\PST@Border(0.1300,0.9680) (0.1300,0.1344) (0.7524,0.1344) (0.7524,0.9680) (0.1300,0.9680)

\catcode`@=12
\fi
\endpspicture